# Empirical study of turning and merging of pedestrian streams in T-junction

J Zhang[1], W Klingsch[1], T Rupprecht[1], A Schadschneider[2] and A Seyfried[3,4]

[1] *Institute for Building Material Technology and Fire Safety Science, Wuppertal, University, Pauluskirchstrasse 11, 42285 Wuppertal, Germany.* [jun.zhang@uni-wuppertal.de](mailto:jun.zhang@uni-wuppertal.de), [klingsch@uni-wuppertal.de](mailto:klingsch@uni-wuppertal.de), [rupprecht@uni-wuppertal.de](mailto:rupprecht@uni-wuppertal.de)

[2] *Institut für Theoretische Physik, Universität zu Köln, 50937 Köln, Germany.* [as@thp.uni-koeln.de](mailto:as@thp.uni-koeln.de)

[3] *Computer Simulations for Fire Safety and Pedestrian Traffic, Wuppertal University, Pauluskirchstrasse 11, 42285 Wuppertal, Germany.* [seyfried@uni-wuppertal.de](mailto:seyfried@uni-wuppertal.de)

[4] *Jülich Supercomputing Centre, Forschungszentrum Jülich GmbH, 52425 Jülich, Germany*

***Abstract:*** *Agent-based models are becoming indispensable tools in building design, safety assessments as well as management of emergency egress. However, reliable calibrations of these models should be mandatory before they are used in practice. To improve the database for model calibration we present results from two experiments at a T-junction and a corner. In such structures the dynamics of pedestrian streams is complex and up to now not studied systematically. To understand it deeply, series of well-controlled laboratory experiments are conducted. The Voronoi method, which is used to analyze the experiments, allows high resolution and small fluctuation in time and space. From the results, it is found that the fundamental diagrams of pedestrian flow in T-junction are not the same before and after merging. At the same density, the velocities of pedestrians before merging are smaller than that after merging. To analyze whether turning or merging of the stream is responsible for this discrepancy, we compare the fundamental diagrams of pedestrian flow in T-junction with the flow at a single corner. The fundamental diagrams of the streams in front and behind the corner agree well and are also in accordance with that from T-junction flow after merging. Besides, space-resolved measurements for the density, velocity profiles are obtained using Voronoi method. These maps offer important information about dangerous spots and thus enable to improve egress management and facility design.*

## 1. Introduction

Plenty of agent-based models for pedestrian dynamics, including continuous models (Chraibi et al., 2010; Helbing et al., 2000) and discrete models (Bandini et al., 2007; Nagatani et al., 2002; Blue & Adler, 2001; Song et al., 2006; Kirchner & Schadschnerider, 2002; Tajima et al., 2001; Tajima & Nagatani, 2002; Peng & Chou, 2011), have been proposed for research in varied contexts. Uni- and bidirectional pedestrian movement in corridors (Nagatani et al., 2002; Blue & Adler, 2001), pedestrian movement from a single-exit room (Helbing et al., 2000; Song et al., 2006; Kirchner & Schadschnerider, 2002), bottleneck flow (Tajima et al., 2001) have been studied with them. Moreover, commercial software like FDS+Evac (Korhonen & Hostikka, 2009), Pedgo (Pedgo, 2005) and Simulex (IES, 2009) et al. have also been developed based on these models. They are becoming indispensable tools in building design, safety assessments as well as management of emergency egress.

In this situation, reliable data for calibration of these models are crucial. To improve the database several laboratory experiments and field studies are carried out to obtain more and better data describing the properties of transport system. From these studies, qualitative phenomena such as lane formation in bidirectional flow (Ma et al., 2010; Hoogendoorn & Daamen, 2004; Kretz et al., 2006), stop-and-go waves (Johansson & Helbing, 2008; Portz & Seyfried, 2010) are observed. Some quantitative data describing pedestrian properties in different conditions are also obtained. Unfortunately, the empirical database is insufficient and has considerable disagreement among different studies and guidelines. Even for the fundamental diagram in simple corridors, the maximum flow as well as the density where the maximum flow appears has a broad range (Schadschneider et al., 2009; Seyfried et al., 2009). For more complex types of facilities, like stairs, crossings, T-junctions or corners the data base is almost not existent. E.g. for T-junctions or corners there are only few studies (Tajima & Nagatani, 2002; Peng & Chou, 2011) dealing with such types of facilities. Nevertheless, T-junctions are important parts of most buildings. In such structures, bottleneck flow, merging flow or split flow could take place. Around corners, the dynamic of pedestrian streams is complex. E.G. we don't know how the effective width of the corridor reduces at corners and how the effective width changes with increasing flow.

In this study, we carried out series of well-controlled experiments in T-junctions and corners to investigate the merging and turning of the streams. The structure of the paper is as follows. In Section 2 we describe the setup of the experiments. The analysis methodology and main results will be exhibited in Section 3. Finally, the conclusions from our investigations will be discussed.

## 2. Experimental setup

Figure 1 shows sketches and two snapshots of the experiment setup. Seven runs of experiments in T-junction and six runs in corners with the same corridor width b = 2.4 m were carried out, respectively. In the T-junction, pedestrian streams move from the left and right branches oppositely and then merge into the main stream. The angle of the corner experiment is $90°$. To regulate the pedestrian density, we set different widths of the entrance, which is 4 m away from the corridor, from 0.5 m to 2.4 m in each run. More than 300 pedestrians participated in the experiments, which makes the duration of the run long enough to get stationary states. The average age and body height of the tested persons was 25 ± 5.7 years and 1.76 ± 0.09 m (range from 1.49 m to 2.01 m), respectively. They mostly consisted of German students of both genders. The free velocity $v_0$ = 1.55 ± 0.18 m/s was obtained by measuring 42 participants' free movement.

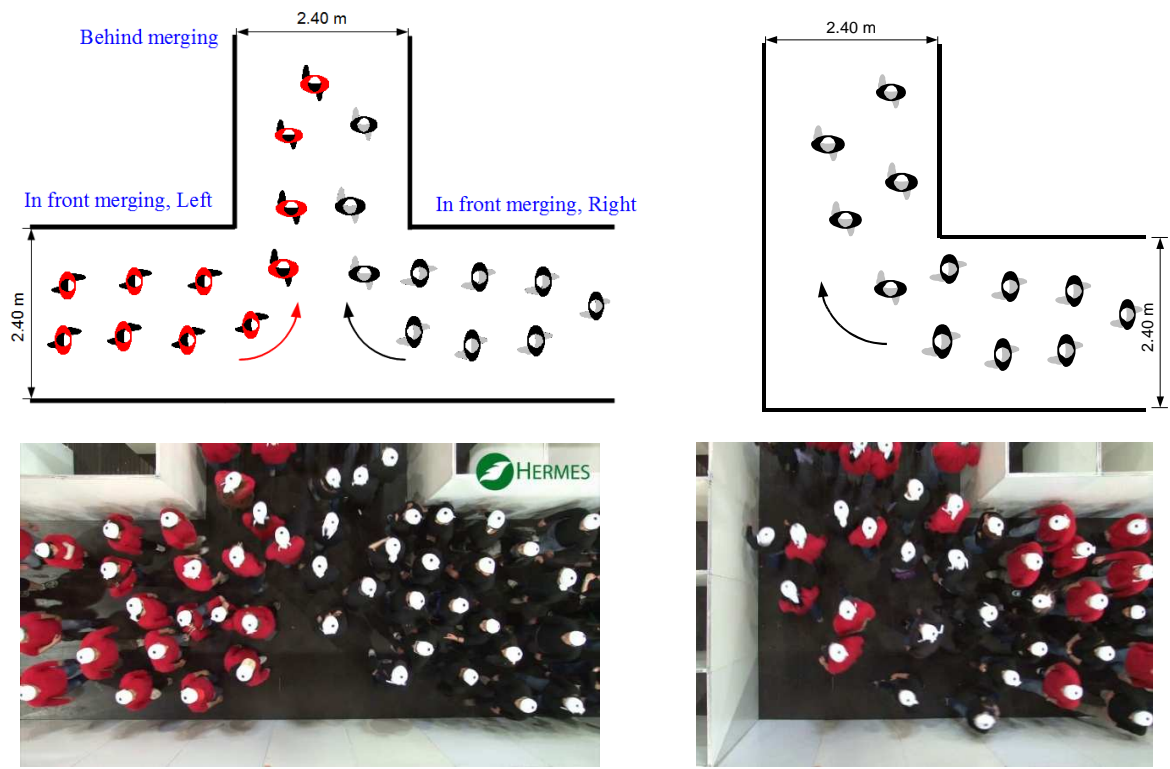

Figure 1: Setup of pedestrian experiments in T-junction and corner.

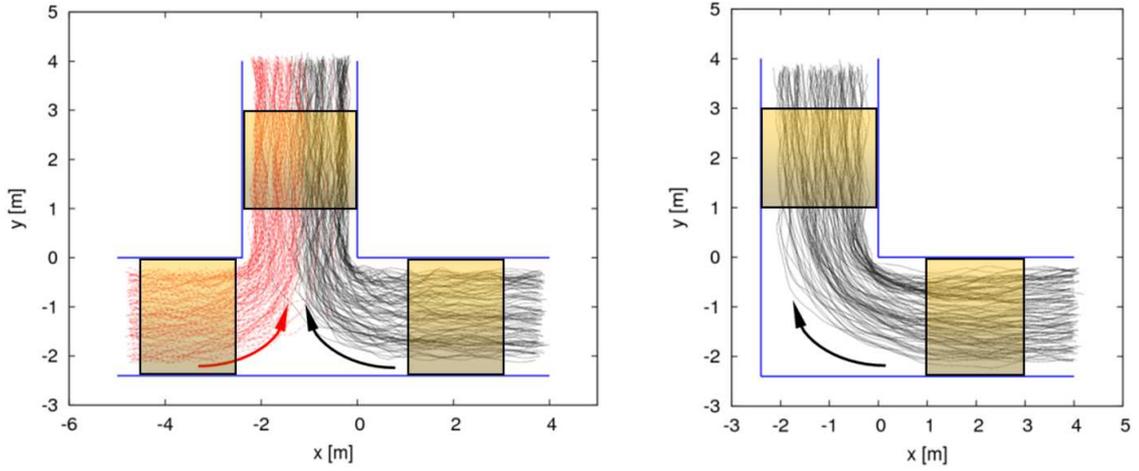

Figure 2: Trajectories from two runs of the experiments extracted by *Petrack*. Yellow rectangles mark the measurement areas (see Table 1 for details).

The runs of the experiments were recorded by two synchronized stereo cameras of type Bumblebee XB3 (manufactured by Point Grey). They were mounted on the rack of the ceiling 7.84 m above the floor with the viewing direction perpendicular to the floor. With this height range, all pedestrians can be seen without occlusion at any time. In the end, accurate pedestrian trajectories were extracted from video recordings using the software *PeTrack* (Boltes et al., 2010) automatically. Figure 2 shows the trajectories from two runs of the experiments. They are very useful not only for model calibrations but also for the analysis of the experiments.

## 3. Voronoi method, analysis and results

In Zhang et al., 2011 and Steffen & Seyfried, 2010, the effect of different measurement methods on the fundamental diagram of pedestrian flow has been compared. This study focuses on the Voronoi method, where the density distribution can be assigned to each pedestrian. This method permits examination on scales smaller than the pedestrians due to its high spatial resolution in combination with low fluctuation.

### 3.1. Voronoi method

At a given time *t*, the Voronoi diagram can be generated from the positions of each pedestrian. It contains a set of Voronoi cells for each pedestrian *i*. The cell area, $A_i$, can be thought as the personal space belonging to each pedestrian *i*. Then the density and velocity distribution over space (see Figure 3) can be defined as

$$\rho_{xy} = 1/A_i \quad \text{and} \quad v_{xy} = v_i(t) \quad \text{if } (x, y) \in A_i \tag{1}$$

Where $v_i(t)$ is the instantaneous velocity of each person (see Zhang et al., 2011).

The Voronoi density and velocity for the measurement area $A_m$ is defined as

$$<\rho>_v (x, y, t) = \frac{\iint \rho_{xy} dx dy}{A_m} \, , \tag{2}$$

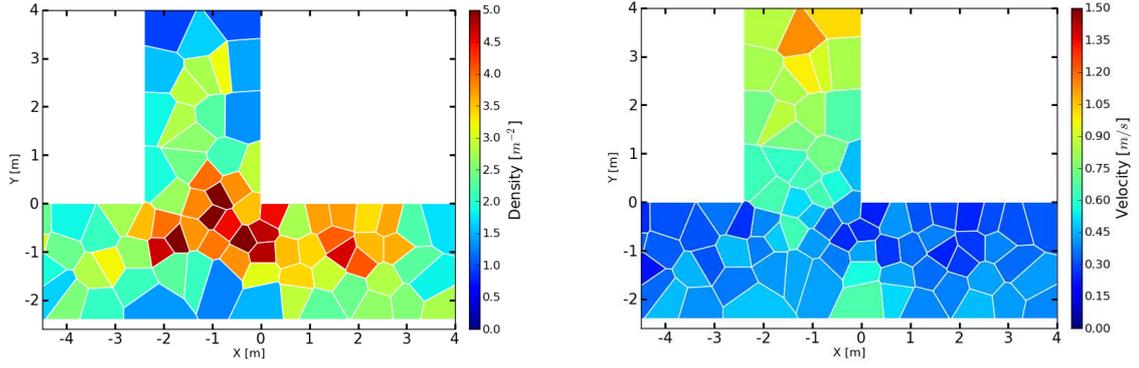

Figure 3: Density and velocity distribution over space obtained from Voronoi method.

$$<v>_v (x,y,t) = \frac{\iint v_{xy} dxdy}{A_m} .$$  (3)

The specific flow

$$<J_s>_v (x,y,t) = <\rho>_v (x,y,t) \cdot <v>_v (x,y,t)$$  (4)

can also be calculated using the Voronoi density and velocity.

### 3.2. Results

In this section, we analyze the pedestrian movement in T-junction and corner based on Voronoi method and the trajectories. The influence of merging and turning streams on the fundamental diagrams are discussed.

### 3.2.1. Fundamental diagram

Firstly, we study the fundamental diagram of the pedestrian flow in T-junction. We choose three different regions, in front merging (left and right) and behind merging, with the same size (4.8 m$^2$) in the geometry as measurement areas, as seen in Table 1. The widths of the measurement area correspond to the width of the corridor and the lengths are 2 m.

Table 1: The locations of measurement areas in the geometry

| Measurement areas | | Range [m] | $A_m$ [m$^2$] |
|---|---|---|---|
| T-junction | In front merging, left | x $\in$ [–4.5,–2.5], y $\in$ [–2.4, 0] | 4.8 |
| | In front merging, right | x $\in$ [1.0, 3.0], y $\in$ [–2.4, 0] | |
| | Behind merging | x $\in$ [–2.4, 0], y $\in$ [1.0, 3.0] | |
| Corner | In front turning | x $\in$ [1.0, 3.0], y $\in$ [–2.4, 0] | |
| | Behind turning | x $\in$ [–2.4, 0], y $\in$ [1.0, 3.0] | |

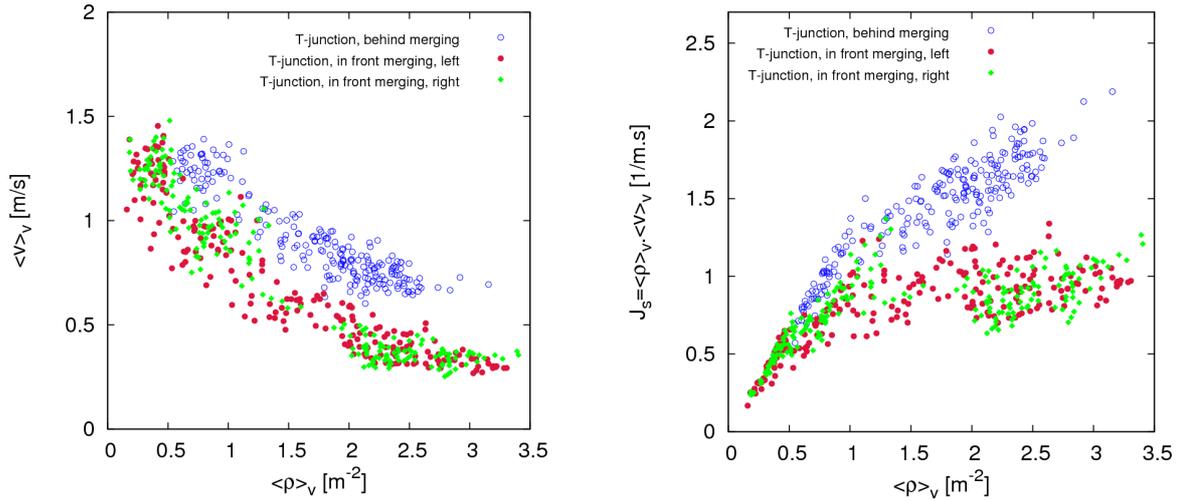

Figure 4: Fundamental diagram of pedestrian stream in T-junction

In Figure 4, we compare the fundamental diagrams obtained from all T-junction experiments. One finds that the fundamental diagrams of the two branches, in front merging, left and right, match well. For densities $\rho > 0.5$ m$^{-2}$, however, at the same density the velocities in front of the merging (at the branches) are significantly lower than that measured behind the merging of the streams. This discrepancy becomes more distinct in the relation between density and specific flow. Behind the merging, the specific flow increases continuously with the density till 2.5 m$^{-2}$. While In front of the merging, the specific flow nearly remains constant for density $\rho$ between 1.5 m$^{-2}$ and 3.5 m$^{-2}$. Thus, there is no unique fundamental diagram describing the relation between velocity and density for the complete system.

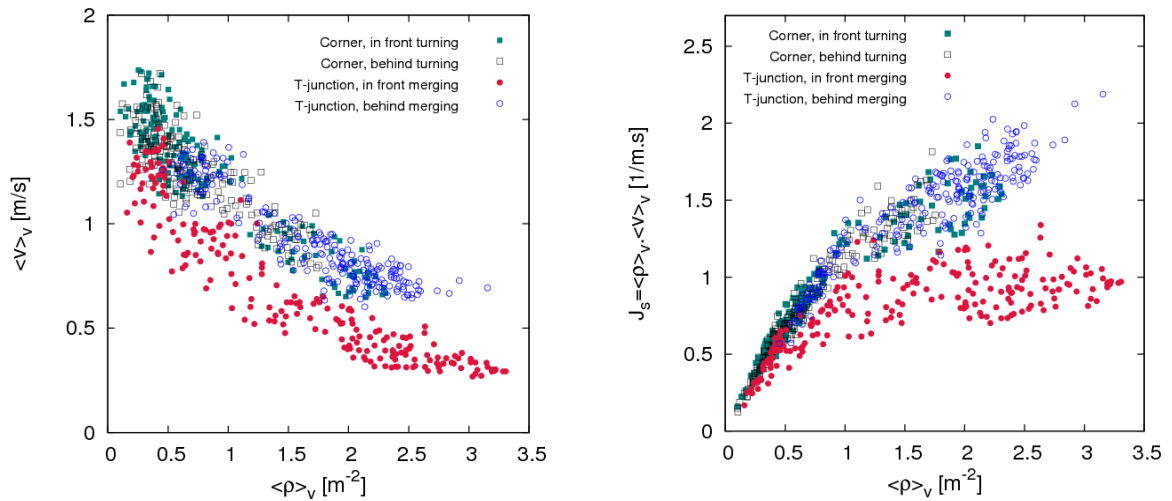

Figure 5: Comparison of fundamental diagrams of pedestrian flow in T-junction and corner

There could be many different causes for this discrepancy. The main processes acting on the pedestrian streams in T-junction are merging and turning. In this study, we compare the T-junction data with data at a corner to check whether the merging or the turning is responsible for this difference. For the comparison, we choose the measurement area with the same size and location as in T-junction. In Figure 5, it can be seen that the fundamental diagrams of streams in front and behind corner agree well and no differences appear. They are also in accordance with that from T-junction flow behind the merging. Thus we assume that the difference of the fundamental diagram in front and behind the merging at a T-junction does not arise from the turning of streams around corner. However, we cannot conclude whether the merging behavior itself or the congestions caused by it lead to the difference. Since the corridor widths are the same for the three parts of our T-junction, as a result, the congestion after merging is obvious and should not be neglected. Another possible reason for this could be behavior of pedestrians. In front of the merging pedestrians standing in a jam and do not perceive where the congestion disperse or whether the jam lasts after the merging. In such a situation, it is questionable whether an urge or a push will lead to a benefit. Thus, an optimal usage of the available space is unimportant and pedestrians could choose other movement pattern.

### 3.2.2. Topographical information for density, velocity and flow

In the last section we have shown that the fundamental diagram can change with the position of the measurement area. To analyze the spatial dependency of density, velocity and flow, we use the Voronoi method to measure these quantities in areas smaller than the size of pedestrians. We calculate the Voronoi density, velocity and specific flow over small regions (10 cm × 10 cm) each frame. Then the spatiotemporal profiles of density ($\overline{\rho}(x,y)$), velocity ($\overline{v}(x,y)$) and specific flow ($\overline{J}_s(x,y)$) can be obtained over the stationary state separately for each run as follows:

$$\overline{\rho}(x,y) = \frac{\int_{t_1}^{t_2} <\rho>_v (x,y,t)dt}{t_2 - t_1} \quad , \tag{5}$$

$$\overline{v}(x,y) = \frac{\int_{t_1}^{t_2} <v>_v (x,y,t)dt}{t_2 - t_1} \quad , \tag{6}$$

$$\overline{J}_s(x,y) = \overline{\rho}(x,y) \cdot \overline{v}(x,y) . \tag{7}$$

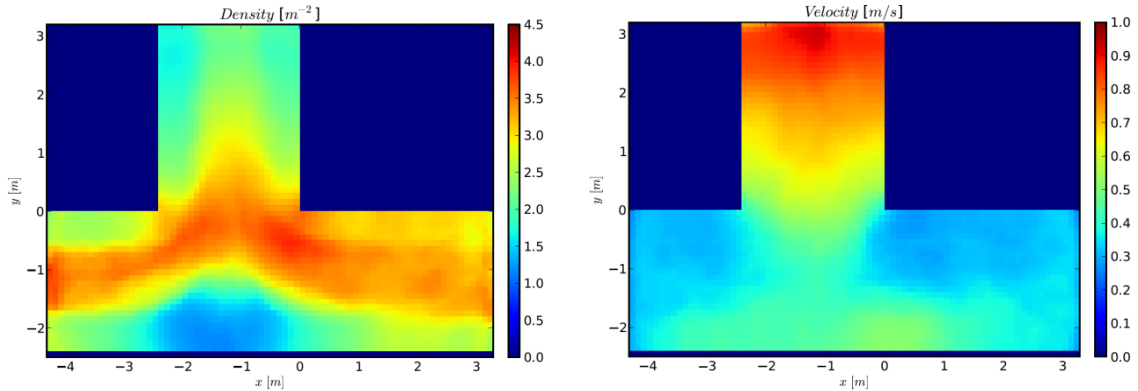

Figure 6: The profiles of density and velocity in T-junction for one run of the experiments.

Figure 6 shows the profiles for one run. These profiles provide new insights into the spatial dynamics of the motion and the sensitivity of the quantities to other potential factors. The density in the T-junction is not homogeneous and a higher density region appears near the junction. The lowest density region is located at a small triangle area, where the left and right branches begin to merge. The densities in the branches (before the merging region) are not uniform and are higher over the inner side, especially near the corners. In other words, pedestrians prefer to move along the shorter and smoother path. Moreover the density profile shows obvious boundary effects. The spatial variation of the velocity is different. From the velocity profile, we can see that pedestrian velocity is nearly homogeneous in the branches. Boundary effect does not occur for the velocity distribution. The velocity becomes larger after pedestrians arriving at the main stream and it increases persistently along the movement direction.

## 4. Summary

In this study, series of well-controlled laboratory pedestrian experiments were performed in T-junctions and corners. The whole processes of the experiments were recorded by two video cameras. The trajectories of each pedestrian are extracted with high accuracy from the video recordings automatically using *PeTrack*. We choose the Voronoi method to analyze the experimental data for its high quality. The fundamental diagrams of pedestrian flow in front and behind the merging in T-junction are compared and discrepancies are observed. To study the causes of these differences, we compare them with data at a corner. In this way, we could test the influence of merging and turning on the pedestrian stream. It is found that the fundamental diagrams of streams in front and behind the turning at the corner agree well and are in accordance with that from T-junction flow behind the merging. The cause for the discrepancies is not the turning of the stream. However, we cannot conclude whether the merging behavior itself or the congestions caused by it lead to the difference at present. The profiles of the density, velocity and specific flow measured with the Voronoi method allow determining critical locations and optimization possibilities. All of these empirical data will be useful for the facility design and model calibrations.